# Photodetection in silicon beyond the band edge with surface states

**T. Baehr-Jones, M. Hochberg**

*Department of Applied Physics, California Institute of Technology, 1200 E California Blvd., Pasadena CA 91125*
*tbaehrjones@yahoo.com*

**A. Scherer**

*Department of Applied Physics, California Institute of Technology, 1200 E California Blvd., Pasadena CA 91125*
*etcher@caltech.edu*

**Abstract**: Silicon is an extremely attractive material platform for integrated optics at telecommunications wavelengths, particularly for integration with CMOS circuits. Developing detectors and electrically pumped lasers at telecom wavelengths are the two main technological hurdles before silicon can become a comprehensive platform for integrated optics. We report on the generation of free carriers in unimplanted SOI ridge waveguides, which we attribute to surface state absorption. By electrically contacting the waveguides, a photodetector with a responsivity of 36 mA/W and quantum efficiency of 2.8% is demonstrated. The photoconductive effect is shown to have minimal falloff at speeds of up to 60 Mhz.

**References and Links**
1. M. Lipson, "Guiding, modulating, and emitting light on silicon - Challenges and opportunities," IEEE Journal of Lightwave Technology **23**, 4222-4238 (2005).
2. B. Jalali, S. Fathpour, "Silicon photonics," IEEE Journal of Lightwave Technology **24**, 4600-4615 (2006).
3. T. Baehr-Jones, M. Hochberg, C. Walker, A. Scherer, "High-Q Resonators in thin silicon-on-insulator," Applied Physics Letters **85**, 3346-3347 (2004).
4. C. P. Michael, M. Borselli, T. J. Johnson, C. Chrystal, O. Painter, "An optical fiber-taper probe for wafer-scale microphotonic device characterization," Optics Express **15**, 4745-4752 (2007).
5. A. Liu, *et al.* "High-speed optical modulation based on carrier depletion in a silicon waveguide," Optics Express **15**, 660-668 (2007).
6. V. Almeida, R. Panepucci and M. Lipson, "Nanotaper for compact mode conversion," Optics Letters **28**, 1302-1304 (2003).
7. A. Liu, R. Jones, O. Cohen, D. Hak, M. Paniccia, "Optical amplification and lasing by stimulated raman scattering in silicon waveguides," IEEE Journal of Lightwave Technology **24**, 1440-1455 (2006).
8. R. A. Soref, "Single-Crystal silicon – a new material for 1.3 and 1.6 mu-m integrated-optical components," Electronics Letters **21**, 953-954 (1985).
9. T. K. Liang, H. K. Tsang, I. E. Day, J. Drake, A. P. Knights, M. Asghari, "Silicon waveguide two-photon absorption detector at 1.5 μm wavelength for autocorrelation measurements," Applied Physics Letters **81**, 1323-1325 (2002).
10. G. Roelkens, D. Van Thourhout, R. Baets, R. Notzel, M. Smit, "Laser emission and photodetection in an InP/InGaAsP layer integrated on and coupled to a Silicon-on-Insulator waveguide circuit," Optics Express **14**, 8154-8159 (2006).
11. J. Liu, D. Pan, S. Jongthammanurak, K. Wanda, L. Kimerling, J. Michel, "Design of monolithically integrated GeSi electroabsorption modulators and photodetectors on an SOI platform," Optics Express **15**, 623-628 (2007).

12. M. W. Geis, S. J. Spector, M. E. Grein, R. T. Schulein, J. U. Yoon, D. M. Lennon, S. Deneault, F. Gan, F. X. Kaertner, T. M. Lyszczarz, "CMOS-Compatible All-Si High-Speed Waveguide Photodiodes With High Responsivity in Near-Infrared Communication Band," IEEE Photonics Technology Letters **19**, 152-154 (2007).
13. M. Borselli, T. Johnson, O. Painter, "Beyond the Rayleigh scattering limit in high-Q silicon microdisks: theory and experiment," Optics Express **13**, 1515-1530 (2005).
14. V. Almeida, R. Panepucci, M. Lipson, "Nanotaper for compact mode conversion," Optics Letters **28**, 1302-1304 (2003).
15. S. Fathpour, K. Tsia, B. Jalali, "Energy harvesting in silicon Raman amplifiers," Applied Physics Letters **89**, (2006).
16. D. Taillaert, *et al.* "An out-of-plane grating coupler for efficient butt-coupling between compact planar waveguides and single-mode fibers," IEEE Journal of Quantum Electronics **38**, 949-955 (2002).
17. M. Casalino, L. Sirleto, L. Moretti, F. Della Corte, I. Rendina, "Design of a silicon resonant cavity enhanced photodetector based on the internal photoemission effect at 1.55 μm," J. Opt. A.: Pure Appl. Opt. **8**, 909-913 (2006).
18. S. M. Sze, *Physics of Semiconductor Devices* (John Wiley & Sons, 1981).
19. T. Baehr-Jones and M. Hochberg Caltech, 1200 East California Blvd, Pasadena, California, 91125, USA are preparing a manuscript entitled "All Optical Modulation in a Silicon Waveguide based on a Single Photon Process."
20. Lide, D. *CRC handbook of chemistry and physics* (CRC Press, 2006).T. Baehr-Jones, *et al.* "Analysis of the tuning sensitivity of silicon-on-insulator optical ring resonators," IEEE Journal of Lightwave Technology **23**, 4215-4221 (2005).
21. Y. Liu, C. W. Chow, W. Y. Cheung, H. K. Tsang, "In-Line Channel Power Monitor Based on Helium Ion Implantation in Silicon-on-Insulator Waveguides." IEEE Photonics Technology Letters **18**, 1882-1884 (2006).

## 1. Introduction

Silicon is an extremely attractive material platform for integrated optics at telecommunications wavelengths [1], particularly for integration with CMOS circuits [2]. Low loss waveguides [3], high-Q resonators [4], high speed modulators [5], efficient couplers [6], and optically pumped lasers [7] have all been demonstrated. Developing detectors and electrically pumped lasers at telecom wavelengths are the two main technological hurdles before silicon can become a comprehensive platform for integrated optics.

Silicon's bandgap of 1.12 eV makes it challenging to build a silicon-based detector in this near infrared. Silicon has minimal absorption of photons in this regime, at least in the bulk, defect-free case [8]. Two-photon absorption can potentially be used to circumvent this limit and build a detector [9], but for practical power levels efficiency is poor. Approaches to detection have typically relied upon bonded III-V materials [10], on integrating Germanium or SiGe [11], or more recently, through volume defect creation via ion implantation [12].

A photoconductive effect has also been observed in undamaged silicon waveguides, and has been attributed to an effect from the surface of the waveguide, though quantum efficiencies of only .24% were shown [12]. Here we show a photodetector based on surface states of a SOI ridge waveguide. Because of the large modal overlap with the surface of the waveguide for our particular geometry, photons are more efficiently absorbed, and a quantum efficiency of 2.8% is obtained.

*1.1 Waveguide geometry*

It is well known that defect states can form at the edge of a crystalline semiconductor. Such defects are known to contribute substantially to the optical losses of silicon waveguides [13]. Most low-loss silicon waveguide geometries involve fairly large silicon waveguides, on the scale of at least .450 μm x .250 μm [14], and often more than 2 μm x .9 μm [15]. We instead use a geometry of .5 μm x .1 μm [3], obtaining losses of around 5 dB/cm. Figure 1 shows a diagram of the waveguide geometry used, as well as information on the mode distribution and an SEM micrograph of the waveguide. Polymers, oxides or air can be used as claddings. Loss comes primarily from three effects: scattering from residual lithographic roughness, absorption from surface states, and absorption from the bulk silicon. These surface states exist precisely where the optical mode field of the waveguide is at its peak intensity, because the electric field is normal to a high dielectric discontinuity. A grating coupler was used to couple light from a standard fiber optic mode pattern into the ridge waveguide [16].

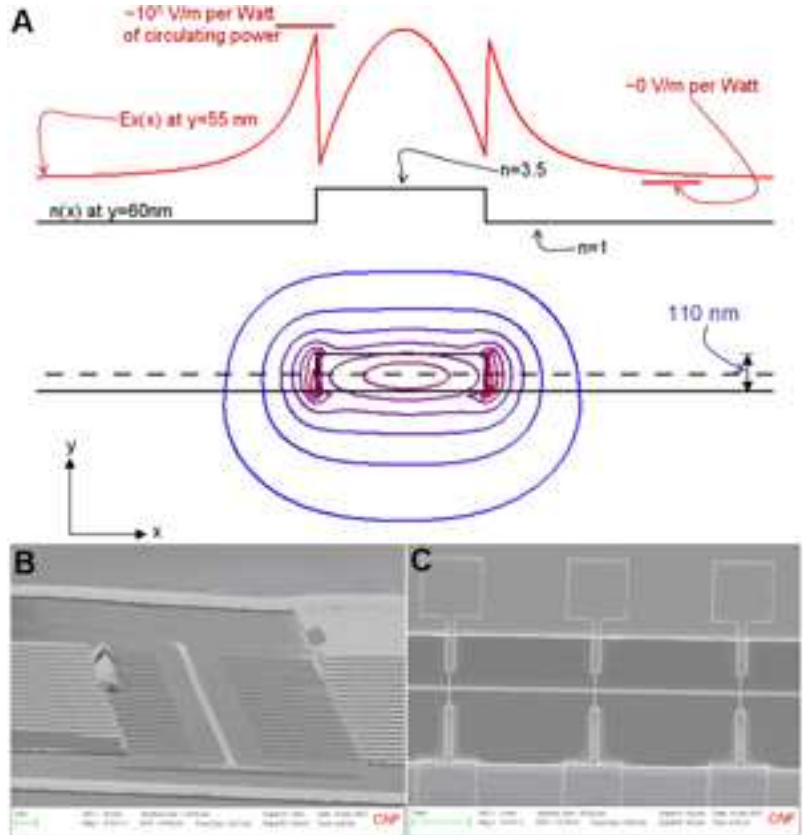

Fig.1. Panel A: a diagram of the waveguide cross section, with the modal pattern for the TE-1 mode overlaid. Contours are drawn in |E| in increments of 10% of max value. For a propagating power of 1 W, the peak electric field will be $10^8$ V/m, In addition, a plot of Ex across the center of the waveguide (dashed line) is shown. Panel B: SEM micrograph of a detector device of type B. C: Another SEM micrograph of a device of type B. A ridge waveguide is contacted by a series of tiny, conductive arms. The optical mode is tightly confined to the ridge waveguide, and does not appreciably touch the metal pads or the surrounding silicon layer.

*1.2 Fabrication*

Devices were fabricated in electronics-grade SOI from Soitec, doped at around $10^{15}$ dopants (Boron)/cm$^3$. No implant or irradiation is performed on the silicon material. The starting material was oxidation thinned to about 110 nm by dry oxidation, singulated into small chips, and patterned using electron-beam lithography on a 100 kV electron beam writer using HSQ resist. The samples were etched with chlorine in an inductively coupled plasma etcher. After removing the residual resist and native oxide, photolithography and evaporation were used to define and deposit aluminum electrodes. No cladding layer was deposited for all devices.

## 2. Electrical measurements

*2.1 Device layout*

The most straightforward way of observing the free-carrier generation effect is to electrically contact the optically active area, and apply a bias voltage. The electron-hole pairs created by the surface state absorption will change the conductivity of the device, and this will result in a photocurrent. Two device types were studied: in type A, the grating couplers were contacted directly with large silicon arms. In type B, small silicon arms can be used to form contacts to the waveguide directly; only around .2 dB of optical insertion loss is endured from .07 um silicon arms. These two device layouts are shown in figure 2. Device type A had a length of .4 mm, while device type B had a length of 1.5 mm, and 40 contacting arms. In both cases, light flows into one grating coupler, through the electrically contacted ridge waveguide, and finally out through another grating coupler. An SEM micrograph of a device of type B is shown in figure 1.

It is important to note that the optical field is separated by tens of microns from the region where the metal pads touch the silicon for both devices. No propagating mode is supported along the tiny conductive arms, and the insertion loss due to each arm is minimal, as confirmed by both simulation and measurement. This is significant, as a Schottky barrier can result in a photoconductive effect for near infrared radiation on the basis of internal photoemission [17]. If the optical mode did touch the metal-silicon barrier, this could be a possible source of the photocurrent that is observed, but the geometry used excludes this possibility.

*2.2 DC I-V Measurements*

The devices were first characterized by coupling a continuous wave laser at 1575 nm through the input coupler, and measuring the DC current that flowed in response to a series of voltages. Several devices were studied, a device of type A (A1) and of two of type B (B1, B2). The layout of this experiment, as well as optical images of the devices are presented in figure 2.

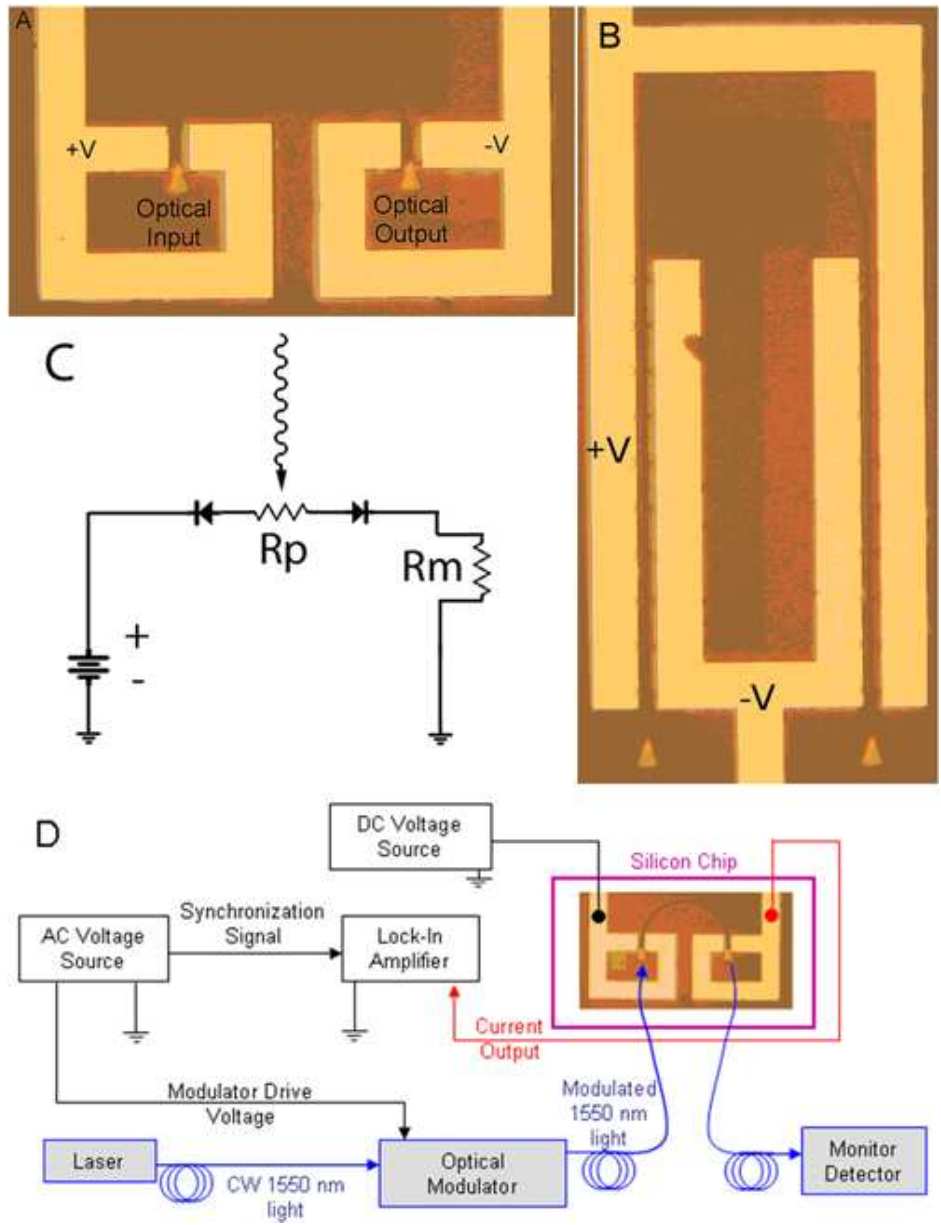

Fig.2. Panel A: A detector of type A. The photoconduction region consists of the loop of the ridge waveguide, while the grating couplers are connected to the metal electrodes. Panel B: A detector of type B. Here the photoconduction region is the intersection of the conduction arms and the waveguide. Panel C: The equivalent circuit, with Rp the photoconductive resistor, and Rm the resistance of the measuring apparatus. The diodes present in the circuit are due to the metal-semiconductor interface. Panel D: A diagram of the entire experimental setup. For DC I-V curves, the lockin would be replaced with a picoammeter.

The measured IV curves for the two devices studied are shown in figure 3. The propagating power shown is the power in the waveguide after the loss due to the grating coupler and other parts of the optical test apparatus. For the data shown, input light at 1575 nm was used. In the case with no incident radiation, the data show the effect of the rectifying contact created by an aluminum electrode on top of lightly p-type silicon [18]. This rectification limits device performance, and decreases the quantum efficiency as a result. Ohmic contacts can easily be formed in future devices with contact doping and annealing. By constraining a heavy p-doping to the pad region only, it should be possible to minimize rectifying behavior, while leave optical behavior unchanged.

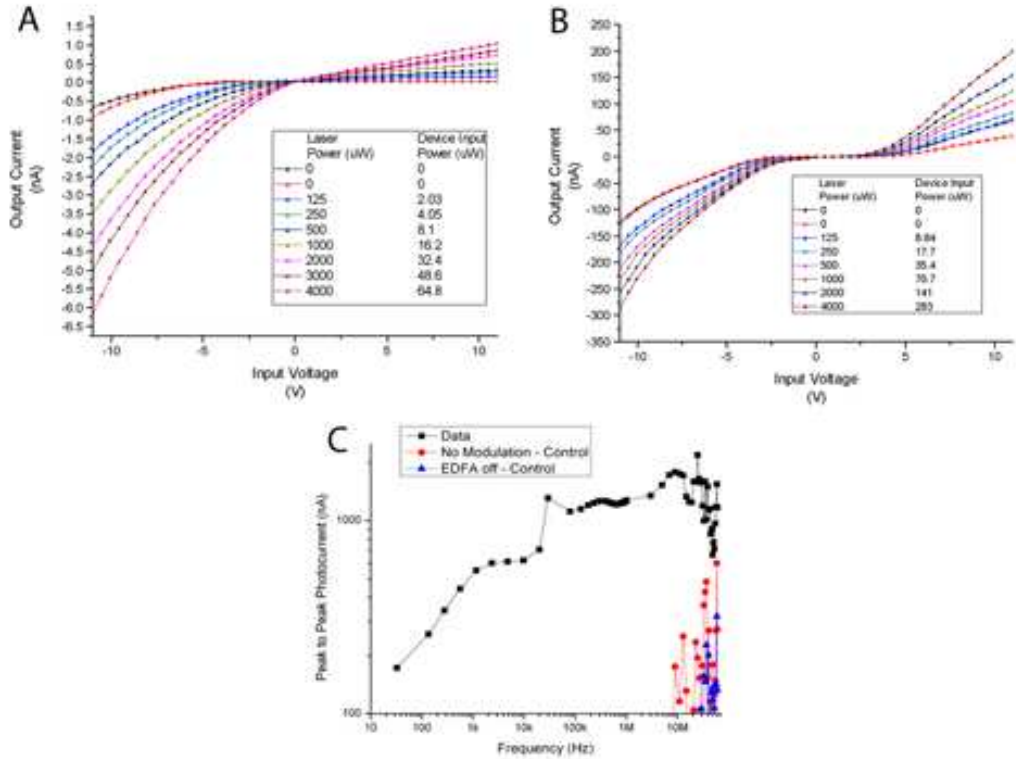

Fig.3. Panel A: DC I-V curves for device A1. Panel B: DC I-V curves for device B1. The power of the test laser used, as well as the propagating laser power in each device is labeled. Slightly different optical paths used in each series of tests result in different levels of propagating power for the same laser power. Current in nA is plotted against voltage in V. Note that panel a shows significantly more rectifying behavior, because there are only two small contacts to the waveguide in this case, while in the case of b there were 40 contacts to the guide. Panel C: The peak to peak output photocurrent of device B2 as a function of frequency. There is minimal change in performance from DC to approximately 60 MHz, where testing was stopped due to limitations of the noise environment where the devices were being tested.

When light is incident on the waveguide, electron hole pairs are generated. These excess electrons and holes are swept towards opposite contacts. This in turn leads to a photocurrent by means of several mechanisms. First, simply the presence of additional carriers in the semiconductor will raise the conductivity. Second, the minority electrons will be able to traverse the reverse biased Schottky diode with relative ease compared to the holes which are the majority carrier in the p-doped silicon.

The current across the device increases along with the laser intensity. For high power, the device response is not linear in laser power, but shows sub-linear behavior; this is attributable to the rectifying contact in series with the photoconductor. In the best case for device A1, the direct current responsivity at an 11 V bias is 0.1 mA/W, while for device B1 the responsivity in the best case is 1.5 mA/W, corresponding to a quantum efficiency of .12%. One of the reasons that device B1's performance is superior is that it has far more semiconductor-metal contacts, allowing more current to flow. It is also possible that the tiny arms connecting the waveguide to the electrodes present additional surface states where optical absorption can occur. Finally, it can be shown that excess electrons and holes can only be cleared from a small portion, around 10 μm of waveguide, from device A1; therefore, devices of type A had much smaller effective lengths than devices of type B. Device B1 was also measured at lower input powers, where performance improved, and the device response became linear.

Device B2 was measured at higher powers, with around 10 mW of input. A responsivity of .12 mA/W at 7 MHz was obtained, with nearly flat performance in frequency; from 1 KHz to 60 MHz, the responsivity changes by only a factor of 3. The slight nonuniformity seen is likely due to electronic parasitics, and perhaps the testing environment. We believe that the device would continue to perform past 60 MHz, however environmental noise precludes measurements at these higher speeds. The lowered responsivity compared to the results for B1 is due to the large amounts of power used in testing at high frequencies.

Further information on the high speed performance of the effect was obtained by all-optical measurements, which are detailed elsewhere [19]. Based on these measurements, the effect involved is extremely fast, and it is likely that with proper supporting electronics, a detector with speeds in the gigahertz could be constructed. The ultimate limit of the device will likely be the time it takes to sweep free carriers across a micron scale silicon device. This speed can easily exceed 10 ghz, as has been shown elsewhere by Geis et al[12]. Their work also, incidentally, suggests that defect absorption centers in silicon can have extremely rapid response times.

These measurements were also performed with similar devices that were clad in PMMA, rather than exposed to the atmosphere. The same photoconductive effect was observed, with very similar performance.

*2.3 Low Intensity Measurements*

It was observed that for larger optical intensities, the dependence of the photocurrent was sublinear, in that doubling the optical input intensity resulted in less than a 2x increase in photocurrent. To better understand the photocurrent process, as well as to estimate the performance in the absence of a rectifying contact, a Princeton Research 5210 lock-in amplifier was used to characterize the photocurrent as a function of optical power. A time constant of 3 s was used for an excitation frequency of 1 KHz, and an input wavelength of 1575 nm. A lithium niobate optical modulator was used to impose a sinusoidal variation on the input intensity, producing a sinusoidally varying photocurrent. The incoming intensity wave is chosen to have nearly full extinction at its lowest point, implying the average power is roughly half of the peak to peak power swing. Figure 4 shows the photocurrent as a function of power for several bias voltages. These measurements were taken for device B1. The responsivity continues to improve for lower optical powers, eventually becoming linear, an effect that is readily explained by the presence of a reverse biased diode, the character of which is seen in the DC curves in figure 3. In the best case, a linear responsivity of 36 mA/W is obtained, corresponding to a quantum efficiency of 2.8% for an 11 V bias. It is worth noting that for higher optical power levels, the responsivity observed with DC measurements is approached.
;

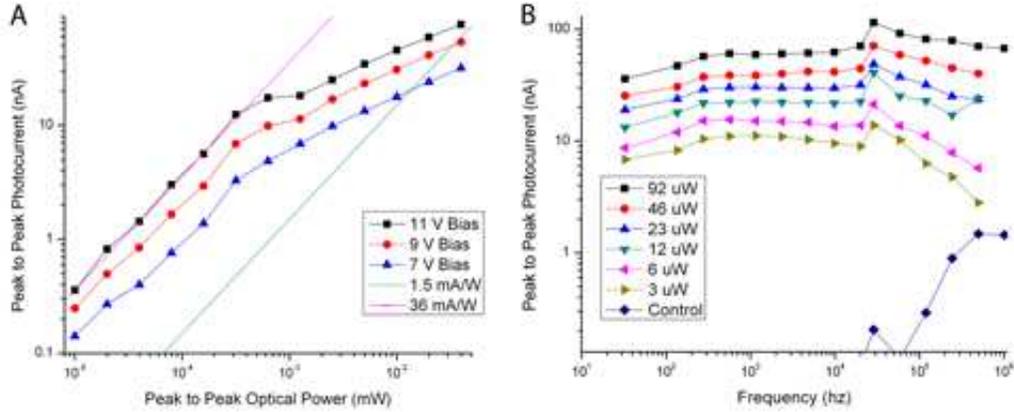

Fig.4. Panel A: Photocurrent as a function of propagating laser power for several bias voltages. The peak to peak photocurrent in nA is reported as well as the peak to peak optical power in the waveguide in mW. The response that would be observed with a perfectly linear 1.5 mA/W and 36 mA/W detector are also shown. Panel B: The photocurrent in peak to peak nA of the device for a 11 V bias voltages and several peak to peak laser powers as a function of frequency up to 1 MHz.

We believe that in the future, more optimal electrode and device geometries will enable the device to demonstrate responsivity at much higher optical powers. This could be achieved on the basis of doping the silicon to increase the conductivity, and by pad implants to decrease the contact resistance.

## 3. Analysis

### 3.1 Exclusion of heating as a significant mechanism

A substantial concern that one might have is that the effect seen is solely due to heating of the waveguide. This would mean that such detector would not be expected function at high speeds, limiting its utility. We note that the conductivity of silicon doped with Boron at $10^{15}$ $1/cm^3$ should not increase until the temperature reaches 230 C or more. This is due to the fact that $n_i$ for silicon does not approach $10^{15}$ $1/cm^3$ until this temperature [18], and thus the conductivity will be mainly due to dopant contributions. We have observed electrical behavior in our devices commensurate with this; increasing the temperature does not typically increase device current. We can then examine a device of type A, and note that for the effect to be explained solely by heating, the volume 20 $\mu m^3$ would have to be heated, by around 200C. The specific heat of silicon is 702 J $kg^{-1}$ $K^{-1}$, and the density is 2.320 g $cm^{-3}$ [20]. This implies that 6.5 nJ would be required to heat the waveguide by this amount.

The device was seen to function, however, with no frequency rolloff to .5 Mhz with 100 µW of input power. At this higher speed the total amount of energy that can be delivered optically in a single cycle is $2x10^{-10}$ J. Even now, there would not be enough energy to heat the waveguide by the large amount needed. But the waveguide used in device type A is identical to the waveguide used in our loss calibration structures. From this, as well as device A's optical performance, the optical loss in the loop is around 5 dB/cm, implying that only around 4.5% of the energy could be absorbed by all of the loss mechanisms combined. Therefore, the amount of energy that could be expected to participate in a heating mechanism is around 1/1000 of the amount that would be needed for the conductivity to change based on a thermal effect.

It is possible, though unlikely, that the structure could first be heated by the dark current induced by the bias voltage, and then reach a very hot temperature at which the heating response to the optical signal was more pronounced than this analysis would suggest. Experiments were performed in which the sample was heated to a number of temperatures greater than 25 C, and the photocurrent observed. Generally, the photocurrent decreased quickly as temperatures rose. In a typical instance, a temperature rise of 20 C decreased the photocurrent by a factor of 2. This suggests that if there is heating from the DC current, the optical response would actually decrease. It also suggests that the possibility of the waveguide being heated to 200 C and this being the source of the photocurrent is even more remote. One explanation for the decreasing photoconductive effect is the effect of temperature on minority carrier lifetimes [18]; the photogenerated electrons will not contribute to the photocurrent if they recombine before reaching an electrode. When cooled to 25C, the original performance was restored on all devices tested.

Another experiment was performed that also suggests that heating cannot be the source of the effect. As is well known [21], thermal heating in a ring resonator results in significant shifts in the resonance peaks. This is due to the fact that silicon's index of refraction will change with temperature. The resonance peaks of a ring resonator on the same chip as the detector devices were scanned using several different optical power levels. The levels ranged from around 10 μW to 1 mW, which in the 1 mW case is more than 10 times the power level used in most of the detection experiments. In all cases, negligible shift in peak appearance or location was observed. However, heating the silicon chip by even 40 C shifted the location of the peaks dramatically. This suggests that if there is any optical heating in our experiments, it is likely to be extremely small, probably less than .1 C. This is orders of magnitude less heating than could possible explain a shift in photocurrent.

*3.2 Exclusion of two-photon absorption as a significant mechanism*

It is well known [9] that for a sufficiently intense beam of optical radiation, silicon can absorb radiation based on two-photon absorption. Given that the dimensions of our waveguides are quite small, and the corresponding intensities are large, this might seem to be a candidate for the effect seen. However, two-photon absorption is a nonlinear process, with the number of photons absorbed depending on the square of the optical intensity. As a result, an effect based on two-photon absorption would be quadratic in input power – that is, the generated current would be quadratic in input intensity. This is already inconsistent with the linear dependence of current with optical power that we have seen. But further experiments were performed.

To determine the amount of two-photon absorption that could be present, calibration structures of two grating couplers and 300 μm of waveguide length were exposed to intense optical radiation beams by using an Erbium-Doped Fiber Amplifier (EDFA). The amount of optical radiation was varied from around 100 μW to 200 mW, and the transmission through the device observed. Linear performance was observed until 100 mW. The nonlinear loss can then be estimated to be at most approximately 250 dB $cm^{-1}$ $W^{-1}$. Measurements with our detectors were performed with 10 μW of power or less, in devices of length 1000 μm. In such a situation, the nonlinear loss stated previously would imply only 0.00025 dB of loss, corresponding to an optical loss of .005%. This would then bound the quantum efficiency to around .0025%, since for every two photons absorbed in this process, a single electron hole pair would be created. This value, however, is nearly a factor of 1000 less than the efficiency observed. Moreover, the nonlinear loss in silicon at high optical powers is not even due solely to two-photon absorption, but mainly to free carriers that have been created by this process5; as a result, the quantum efficiency that could be obtained by two-photon absorption is probably much lower than this.

*3.3 Identification of surface states as free-carrier source*

If one accepts that the photocurrent is caused by the generation of an electron and a hole through absorption at a defect state, then the question of whether these defects are in bulk or are at the surfaces remains open. To this end, it is useful to compare our device to previous work on optical channel monitors that function based on defects from implant damage. The most salient basis for this comparison is the amount of photocurrent per unit waveguide length. Liu et al[21] have build a volume defect photodetector in silicon which achieves a quantum efficiency of 5% with around $2\times10^{15}$ $1/cm^3$ of He implants, but in a device that is 1.7 cm long. Our device achieves nearly the same efficiency in around 1/20 of the length. If we assume that there is one defect per Boron dopant in our device, and one defect per helium implant in Liu's device, then a substantial discrepancy exists. That is, with only half of the defects per unit volume, we achieve a similar quantum efficiency to Liu et al in a device that is 20 times shorter. Other considerations only skew this comparison further; it is likely that there are, on average, multiple defects associated with a Helium ion implanted at 800 KeV as compared to including a dopant during the wafer manufacturing process. Moreover, our waveguide geometry has a substantial portion of the mode outside the waveguide, which would only serve to lower the influence of bulk defects.

On the other hand, both Liu et al [21] and Geis et al [12] both observed at least some photocurrent with undamaged silicon control samples, though much lower quantum efficiencies of around .1% were obtained. Geis et al attributed this, without discussion or supporting data, to a process that occurred on the waveguide surface. We believe that they may have also observed surface state based absorption and generation of electron-hole pairs. The relative difference in efficiencies is probably due to the very different surface modal overlap that our waveguide exhibits.

The best basis for understanding the source of the photoconductive process is an analysis of the source of waveguide loss; clearly, the generation of an electron-hole pair must correspond to the absorption of a photon and thus a certain amount of optical loss. Borselli et al have recently shown that bulk silicon waveguide losses are no more than .13 dB/cm in p-type SOI [22]. The SOI used was from SOITEC, and was doped with Boron in nearly identical concentrations to the wafers used for our devices. The fractional decrease in power due to bulk silicon loss over the course of a 1.5 mm device is only .45%, which would then be a strict upper bound on the quantum efficiency. Our devices exceed this limit by nearly an order of magnitude. This also strongly suggests that the defect states used by the photoconduction effect must exist at the surface of the silicon waveguide, and are due to the etched silicon facets.

*3.4 Calculation of the surface state density*

The strength of the surface state absorption effect can be characterized by a surface state density, $\sigma$ in units of watts$^{-1}$s$^{-1}$cm$^{-1}$. It identifies the product of the number of surface states and the optical absorption probability per unit of waveguide. The number of electron-hole pairs generated per second per cm of waveguide can be written as:

$$\frac{EHP}{s \cdot cm} = \sigma I \quad (1)$$

Assuming a nearly uniform optical intensity throughout the detector, the responsivity for a device of length L can then be written:

$$R = q\sigma L \tag{2}$$

Here q is the charge of an electron. Based on the responsivity of 36 mA/W for a device of length 1.5 mm, σ can be estimated at $1.5 \times 10^{18}$ watts-1s-1cm-1. This value is comparable to the surface state density calculated for similar samples with all optical measurements [19].

The waveguide loss can also be written as a function of the surface state density.

$$\alpha >= h\nu\sigma \tag{3}$$

Here h is Planck's constant, and ν is the optical frequency. The surface state density estimated above corresponds to a waveguide loss of .8 dB/cm. This is a notable fraction of the waveguide loss of 5 dB/cm that these waveguides typically exhibit.

## 4. Conclusion

Further work needs to be done in order to determine the precise mechanism by which electron-hole pairs are generated. This will determine the ultimate bandwidth limit of the effect, as well as suggest methods for increasing its strength. We believe that with further optimizations of electrode geometries and processing parameters, much better responsivity can be obtained from surface-state absorption, at higher power and speeds. It is likely that the effect could be used to build a photodetector at 1 ghz or higher, which would be useful for commercial telecommunications applications. Our results also serve as additional evidence that surface-state absorption is an important component of waveguide losses in nano-scale silicon waveguides. Measurement of the photoconductivity of waveguide samples could become a useful tool in loss optimization. More generally, our results suggest that novel material functionality can sometimes be obtained by using nanoscale lithography to access and enhance surface effects.

**Acknowledgements**

This work was supported in part by the Air Force Office of Scientific Research and the Cornell Nanoscale Facility, which is funded as part of the National Nanotechnology Infrastructure Network by the National Science Foundation.